# The CoRoT satellite in flight : description and performance.

Auvergne M.[1], Bodin P.[2], Boisnard L.[2], Buey J-T.[1], Chaintreuil S.[1], Epstein G.[1], Jouret M.[2], Lam-Trong T.[2], Levacher P.[3], Magnan A.[3], Perez R.[2], Plasson P.[1], Plesseria J.[11], Peter G.[4], Steller M.[10], Tiphène D.[1], Baglin A.[1], Agogué P.[2], Appourchaux T.[5], Barbet D.[5], Beaufort T.[9], Bellenger R.[1], Berlin R.[4], Bernardi P.[1], Blouin D.[3], Boumier P.,[5], Bonneau F.[2], Briet R.[2], Butler B.[9], Cautain R.[3], Chiavassa F.[2], Costes V.[2], Cuvilho J.[12], Cunha-Parro V.[1] De Oliveira Fialho F.[1], Decaudin M.[5], Defise J-M.[11], Djalal S.[2], Docclo A.[1], Drummond R.,[13], Dupuis O.[1], Exil G.[1], Fauré C.[2], Gaboriaud A.[2], Gamet P.[2], Gavalda P.[2], Grolleau E.[1], Gueguen L.[1], Guivarc'h V.[1], Guterman P.[3], Hasiba J.[10], Huntzinger G.[1], Hustaix H.[2], Imbert C.[2], Jeanville G.[1], Johlander B.[9], Jorda L.[3], Journoud P.[1], Karioty F.[1], Kerjean L.[2], Lafond L.[2], Lapeyrere V.[1], Larqué T.[2], Laudet P.[2], Le Merrer J.[3], Leporati L.[3], Leruyet B.[1], Levieuge B.[1], Llebaria A.[3], Martin L.[3], Mazy E.[11], Mesnager J-M.[2], Michel J-P.[1], Moalic J-P.[5], Monjoin W.[1], Naudet D.[1], Neukirchner S.[10], Nguyen-Kim K.[5], Ollivier M.[5], Orcesi J-L.[5], Ottacher H.[10], Oulali A.[1], Parisot J.[1], Perruchot S.[3], Piacentino A.[1], Pinheiro da Silva L.[1], Platzer J.[1], Pontet B.[2], Pradines A.[2], Quentin C.[3], Rohbeck U.[8], Rolland G.[2], Rollenhagen F.[4], Romagnan R.[1], Russ N.[4], Samadi R.[1], Schmidt R.[1], Schwartz N.[1], Sebbag I.[2], Smit H.[9], Sunter W.[9], Tello M.[2], Toulouse P.[2], Ulmer B.[7], Vandermarcq O.[2], Vergnault E.[2], Wallner R.[10], Waultier G.[3], Zanatta P.[1],[*]

(Affiliations can be found after the references)

**ABSTRACT**

*Context.* CoRoT is a space telescope dedicated to stellar seismology and the search for extrasolar planets. The mission is led by CNES in association with French laboratories and has a large international participation: the European Space Agency (ESA), Austria, Belgium and Germany contribute to the payload, and Spain and Brazil contribute to the ground segment. Development of the spacecraft, which is based on a PROTEUS low earth orbit recurrent platform, commenced in October 2000 and the satellite was launched on December 27[th] 2006
*Aims.* The instrument and platform characteristics prior to launch have been described in ESA publication (SP-1306) . In the present paper we detail the behaviour in flight, based on raw and corrected data.
*Methods.* Five runs have been completed since January 2007. The data used here are essentially those acquired during the commissioning phase and from a long run which lasted 146 days, these enable us to give a complete overview of the instrument and platform behaviour for all environmental conditions. The ground based data processing is not described in detail, the most important method being published elsewhere.
*Results.* It is shown that the performance specifications are easily satisfied when the environmental conditions are favourable. Most of the perturbations, and consequently data corrections, are related to Low Earth Orbit (LEO) perturbations: high energy particles inside the South Atlantic Anomaly (SAA), eclipses and temperature variations, and line of sight fluctuations due to the attitude control system. Straylight due to the reflected light from the earth, which is controlled by the telescope and baffle design, appears to be negligible.

## 1. Introduction

Physical insight into stellar internal structure depends on our ability to compare and adjust models to reproduce the observations. However the number of unknown quantities is larger than the number of measured quantities and many uncertainties remain, preventing accurate tests of the hypotheses contained in the modeling process. Asteroseismology, i.e. the detection of eigenmodes of oscillations, provides new observables (frequencies and amplitudes of eigenmodes) with very high accuracy, and increases the number of constraints on models. It has already proven to be a powerful tool to probe the interior of the Sun. Observations of other stars are needed to test different physical conditions and discriminate among physical processes. Seismology measurements are performed in the bandwidth 0.1 to 10 mHz, covering both pressure and gravity modes of stars. For stochastically excited modes, the frequency measurement precision depends on the length of the observation window, on the mode lifetime, on the amplitude and on the signal-to-noise ratio in the Fourier spectrum. We expect mode lifetimes of about 5 days and amplitudes greater than 2 ppm; the target stars are therefore observed for 150 days with a minimum signal-to-noise ratio of 15 (in terms of power spectral density) to yield a precision on frequency measurements of 0.1 $\mu$Hz. The signal-to-noise ratio should be reached in the frequency interval if the noise level remains of the order of 0.6 ppm in the Fourier spectrum in 5 days of observation for a 5.7 V magnitude star.

With high precision stellar photometry we can detect extrasolar planets by the transit method, measuring the decrease of the stellar flux when the star, the planet and the instrument are almost aligned. The flux decrease is given by :

$$\frac{\Delta F}{F} \propto \left(\frac{R_p}{R_s}\right)^2 \quad (1)$$

---

[*] The CoRoT space mission, launched on December 27th 2006, has been developed and is operated by CNES, with contributions from Austria, Belgium, Brazil , ESA, Germany and Spain. Four French laboratories associated with the CNRS (LESIA,LAM, IAS ,OMP) collaborate with CNES on the satellite development. The authors are grateful to Ian Roxburgh for a careful reading of the manuscript.

where $R_p$ is the radius of the planet and $R_s$ the radius of the parent star. If the impact parameter is equal to 0, the transit duration is :

$$tr = \frac{P}{\pi}\left(\frac{R_s}{a}\right) \qquad (2)$$

where P is the orbital period of the planet and $a$ the radius of the orbit supposed to be almost circular.

In addition to hundreds of Uranus-like or Jupiter-like planets, several telluric planets should be detected, if our hypotheses about accretion models and planets occurrence are correct. To reach this goal the required photometric precision is $7.10^{-4}$ for a V=15.5 magnitude star for one hour integration. Up to 12000 target stars, in a field of view of 4 square degrees, are simultaneously observed during a 150-day period. With at least five different fields of view, more than 60 000 stars will be followed during 150 days over the whole mission. Between each 150 days runs a short run of typically 25 days is performed.

The layout of our paper is as follow. In section 2 we recap the mission status and the spacecraft characteristics. Then the environmental conditions for a Low Earth Orbit (LEO) and the impact on the photometric data are recalled in section 3. In addition some results obtained during the commissioning phase are presented. Section 4 briefly describes the payload architecture and gives details on the payload and satellite behaviour. We develop important features of some functional subsystems such as the attitude control system, the temperature control system and the detection chain. On board and ground based data processing are summarised in sections 5 and 6 and photometric performances are presented in section 7 for the AsteroSeismology (AS) and Planet Finder (PF) channels. In section 8 a brief summary closes the paper.

## 2. The satellite.

The satellite was launched successfully on 27/12/2006 into a polar orbit. The orbit is so close to the target one that no orbit corrections were necessary. The inclination of the orbit plane, which is 90.002 degrees (for a target inclination of 89.184 degrees), produces a drift of the orbit plane of one degree per year toward increasing right ascension (RA). The apogee and perigee are respectively of 911 and 888 km. The orbital period is 6184 seconds. Up to April 2008 five observing runs have been successfully completed. Their durations are of 59, 28, 157, 146 and 20 days, for respectively the IRa01, SRc01, LRc01, LRa01 and SRa01 (LR stands for Long Run, SR for Short Run, IR for Initial Run and the small letters a and c for galactic anti-centre or galactic centre direction). For a complete overview of the mission see for example the CoRoT Book, 2006 and Boisnard & Auvergne, 2004.

The spacecraft is based on a generic PROTEUS platform (Landiech and Douillet, 2004) developed under a close collaboration between the Centre National d'Etudes Spatiales (CNES) and Alcatel Space Industry (now re-named Thales-Alenia Space). It consists not only of a generic platform but also of a generic ground segment. The platform was designed for Low Earth Orbit (LEO) missions with a maximum altitude of 1300 km. Its lifetime is 3 years and elements sensitive to aging or radiation are sized to 5 years. The satellite is operated from the CoRoT Control Centre located in Toulouse, sharing the facilities of the PROTEUS satellite family. The preparation of the observing runs, the payload command control and the pre-processing of the scientific data are achieved by the CoRoT Mission Centre in collaboration with participating laboratories. A network of CNES ground stations (Kiruna, Aussaguel, Hartebeesthoek, Kourou), and one mission specific secondary ground station located in Alcantara (Brazil), are used for communication with the satellite and/or reception of technical and scientific telemetry.

The interfaces for a dedicated mission are at satellite level between the platform and the payload, and at ground level between the Control Centre and the specific Mission Centre. In the configuration adopted for CoRoT, several sub-systems have been upgraded : Li-Ion battery, high capacity magneto torquer bars and new star trackers (SED-16). The total weight is 626 kg, including 300 kg for the payload. A mass memory provides at end of life a storage capacity of 2 Gbits for the payload and housekeeping data. The power is provided by two solar arrays which feed directly to the battery giving a non regulated voltage in the interval $23 - 37$ V. and a power of 1 kW. The battery management is driven by autonomous software. An important functional chain is the Attitude Control System (ACS). At platform level the pointing is provided by star trackers, inertial wheels, magneto-torquers and gyrometers, giving an angular stability of 16 arc seconds. Since CoRoT's requirement is about 0.5 arcsec rms, to reach this value it was necessary to include in the control loop ecartometry computations based on the position of two stars on the payload focal plane.

## 3. Environmental conditions.

As the orbit is at an altitude close to 900 km the Earth has an influence on the satellite and introduces perturbations on three characteristic time scales; the orbit (and harmonics), the day (and its first harmonic) and the seasons. The main perturbations are:

– Eclipses: the transition light/penumbra/shadow and the reverse produces temperature variations, vibrations in the solar panels at the transition time, and voltage fluctuations when the power supply switches from the solar arrays to the battery (see figure 1).

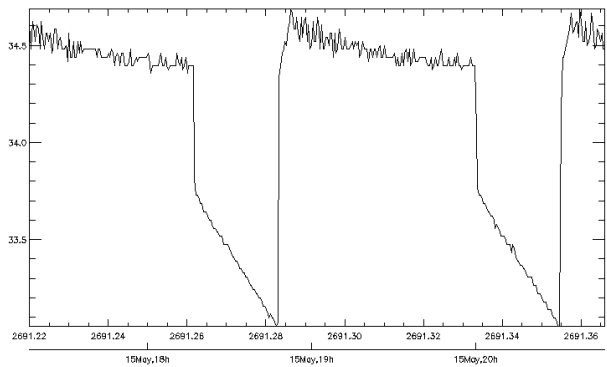

**Fig. 1.** Power supply voltage variations at eclipse ingress/egress on two orbits. Ordinate unit: Volts. Abscissa unit: Julian days. Note that for all plots the Julian origin is January 1 2000., JD0 = 2451545.

– Gravity field and Earth's magnetic field: these induce perturbation torques on the satellite perturbing the satellite attitude.
– South Atlantic Anomaly (SAA): the magnetic field is not a perfect dipole, and the radiation belts are distorted above South America. This is due to the fact that the centre of Earth's magnetic field is offset by 450 km from the geographic centre. Trapped particles are essentially protons with

- energies in the interval 10 keV to 300 MeV. The electrons belts around the pole do not perturb the detectors thanks to a 10 mm aluminium shielding.
- Sun and Earth infrared emissivity: The Sun's position produces seasonal temperature variations, mainly through the focal box radiator. The Earth's infrared emission produces temperature variations on orbit and day time scales.
- Earth albedo: The sunlight is reflected (or not) by the Earth depending on the region overflown, day/night region, cloudy zone and ocean.
- Objects in LEO: At altitudes between 400 and 1500 km there are several thousands of satellites and debris of size greater than 10 cm. When such an object is in the CoRoT field of view (FoV) it can produce a local or global perturbation everywhere in the FoV.

### 3.1. Eclipses.

The eclipse durations vary along a run. At the beginning or at the end of a run (just after or before the equinoxes) the eclipses are longer. Near the solstices there are no eclipses for more than a month (see figure 2 ). The longest eclipse duration is 34.72 mn. All others effects are detailed in the following.

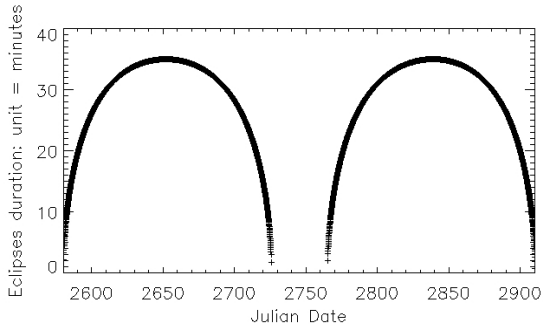

**Fig. 2.** Eclipse durations during the year 2007.

### 3.2. SAA

The radiation flux, intensity of impact and the number of pixels perturbed have been studied during the commissioning period on images specifically acquired for this task. The detection method of impact in the image and results are detailed in Pinheiro da Silva et al., 2008. We recall in the following the main results. Comparisons with theoretical models show that protons are the dominant contribution, the measured flux is in good agreement with the predicted one (see figure 3). The average imparted energy per impact (which should not be confused with the proton energy) is 5160 $e^-$, the most violent event generates more than 1.1 M$e^-$. The mean number of pixels disturbed in an impact is 7.9. It is also found that events with low energy are dominated by secondary particles generated by protons in the surrounding material of the detectors. We define the limits of the SAA such that the probability to have an impact inside a stellar image is less than 50 %. This corresponds to a proton flux of 200 $p^+/cm^2/s$ on the AS channel and 50 $p^+/cm^2/s$ on the PF channel. With this definition the percentage of time spent by the satellite inside the SAA is 7 %. It is a significant contribution to the duty cycle specification which was 90 %. The measured proton flux mapped onto the Earth shows the location of the SSA as in figure 4.

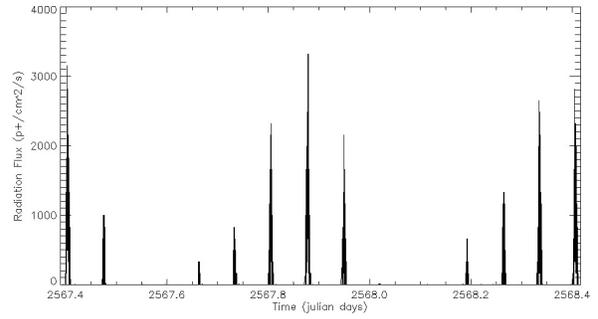

**Fig. 3.** Temporal evolution of the proton flux at the CoRoT altitude.

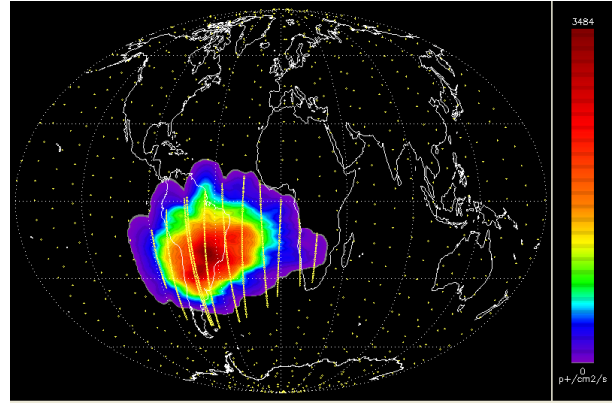

**Fig. 4.** Radiation flux $p^+/cm^2/s$ map. The yellow points are the moment of image acquisition.The oscillations on the edges are due to poor time sampling.

With transient events, proton impacts also produce permanent hot pixels due mainly to atomic displacements in the silicon lattice (Srour et al., 2003, Hopkinson, 1996). The intensity of a bright pixel is not stable in time. On short time scales (few minutes to few hours) rearrangement phase induces abrupt or exponential decrease of the intensity. A long term annealing often follows, and a bright pixel can disappear after several days to a few years. Figure 5 shows one example of the time behaviour of a hot pixel.

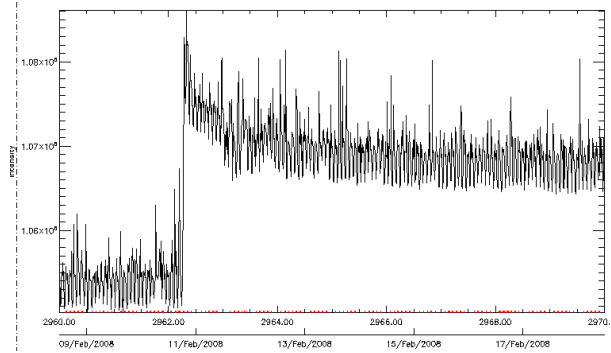

**Fig. 5.** Example of a background raw data light curve (LC) perturbed by a new hot pixel appearing during the crossing of the SAA. In this example the hot pixel intensity decreases exponentially on a 1 day time scale and reaches a stable but higher mean value. Orbital straylight oscillations are also clearly seen.

At the beginning of each run the bright pixels are detected on PF CCDs images as a function of their intensity. The results are shown in figure 6 and in table 1. The evolution of the number of hot pixels appears to be linear in time so that we can extrapolate this number to the satellite end of life.

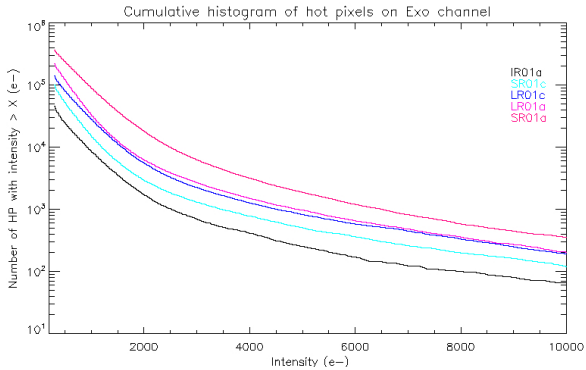

**Fig. 6.** Number of hot pixels having an intensity larger than *x* electrons at the beginning of the five first runs on the PF CCDs.
hot pixel

## 4. Instrument description.

### 4.1. Optics: The telescope and the focal plane.

#### 4.1.1. The telescope.

The optical architecture was driven by the need to minimise straylight coming from the Earth and to provide a field of view of 2.7 × 3.05 degrees. This is shown in figure 7. An off-axis afocal telescope with a 2 stage baffle and a camera made of a dioptric objective and a focal box is the adopted solution (Viard et al., 2006 ). It has the following advantages: a real exit pupil and a real field stop, no central obscuration and a cavity baffle completing the first stage. The baffling architecture is given in figure 8. The collecting surface is 590 cm$^2$. With this collecting surface targets with *V* magnitudes between 5.4 and 9.2 on the AS CCDs and with *R* magnitudes between 11.5 and 16 on PF CCDs, can be observed with a reasonable signal to noise ratio and without saturation of the detectors. The real pupil is at the dioptric objective entrance. To make the pupil surface variations negligible, the temperature of the pupil is stabilised. The residual orbital amplitude is less than 0.05 K. The corresponding relative surface variation (0.05 10$^{-6}$) can be disregarded.

#### 4.1.2. The focal box.

The focal box contains the four CCDs. Detectors are protected against radiation by a 10 mm thick aluminium shielding. The AS CCDs are defocused by 760 $\mu$m toward the dioptric objective to avoid saturation of the brightest stars. A prism in front the PF CCDs gives a small spectrum. The prism dispersion law is non-linear, dispersing more strongly the blue than the red. The PF CCDs plane is close to the best image plane, but as the best focused image depends strongly on the wavelength, we have chosen to focus the blue in order to compensate for the strong blue dispersion (see figure 9). Evaluation of the focal plane position after the launch shows that the CCDs are defocused by 100 $\mu$m more, giving a slightly larger PSF, without any consequences for the photometric precision. The focal box is thermally linked to an external radiator in order to keep the CCDs temperature around −40 C (see section 4.4).

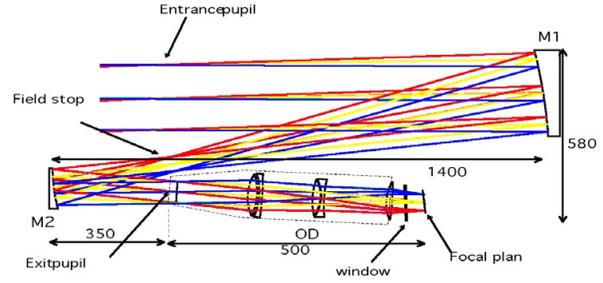

**Fig. 7.** Optical layout.

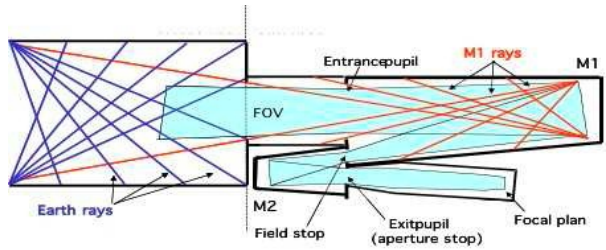

**Fig. 8.** Baffle achitecture.

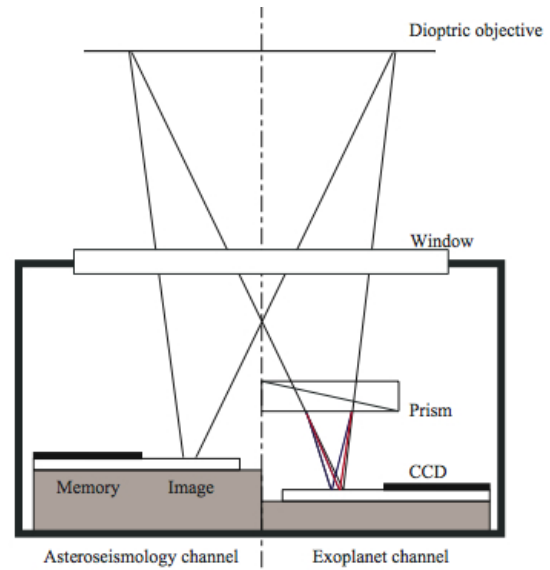

**Fig. 9.** Layout of the focal box.

### 4.2. Optical Characteristics.

#### 4.2.1. Baffle efficiency and straylight.

The background flux is composed of several components: zodiacal light, dark current and straylight from the Earth. The straylight is responsible for the background orbital variations. Ten windows of 50 × 50 pixels each on the AS CCDs and 400 windows of 10 × 10 pixels on the PF CCDs give continuous mea-surements with an integration time of 1, 32, and 512 seconds. The measurement precision is given by the relation:

$$\frac{\sigma_B}{B} = \sqrt{\frac{1}{kB} + \frac{\sigma_r^2}{k\tau B^2}} \qquad (3)$$

where $B$ is the mean background per pixel per second, $\sigma_r$ the readout noise, $\tau$ the binning window rate and $k$ the number of pixels in the window. The binning rate is typically equal to 25 in seismology windows and 1 (no binning) in the PF background windows. With a readout noise of 8 e$^-$, an average background of 20 e$^-$ s$^{-1}$ pixel$^{-1}$, the precision is respectively 4 % and 0.4 % on the AS and PF CCDs. For all runs the background orbital amplitude is less than 0.6 e$^-$ s$^{-1}$ pixel$^{-1}$ (figure 10).

**Table 1.** Number of bright pixels at the beginning of the four runs.

| Date | > 300 e$^-$ | > 1000 e$^-$ | > 10000 e$^-$ |
|---|---|---|---|
| Launch + 35 days | 22314 (0.53%) | 4280 (0.1%) | 33 (0.00079%) |
| Launch + 105 days | 49026 (01.17%) | 7479 (0.18%) | 57 (0.0014%) |
| Launch + 137 days | 69275 (1.65%) | 13865 (0.33%) | 92 (0.0022%) |
| Launch + 294 days | 108435 (2.58%) | 16734 (0.4%) | 101 (0.0024%) |
| Launch + 434 days | 173562 (8.27%) | 43440 (1.03%) | 174 (0.0041%) |
| End of mission (2.5 years) | 8.5% | 1.08% | |

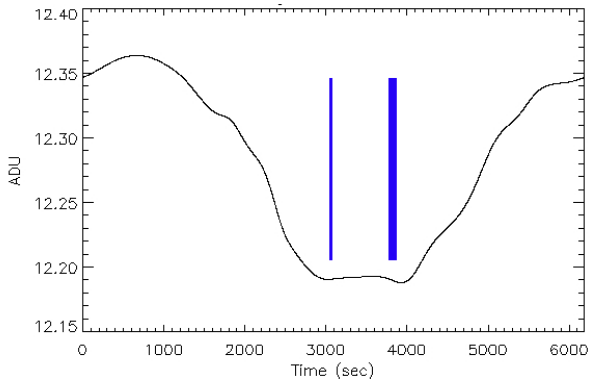

**Fig. 10.** Example of the background orbital variations measured on AS CCDs. Data has been folded on the orbital period. The left blue vertical bar shows the eclipse ingress and the right one the eclipse egress. The peak to peak amplitude is 0.2 ADU pixel$^{-1}$ or 0.4 e$^-$ pixel$^{-1}$

#### 4.2.2. The PSF on the two channels.

Stellar images are close to simulated ones, the diameter of AS PSF is 2 pixels larger than expected (figure 11); the PF CCDs plane being close to best focus the PSF is almost unchanged (figure 12).

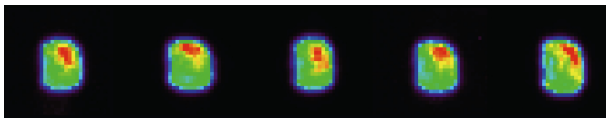

**Fig. 11.** Five AS PSFs. From left to right the positions on CCD A2 are: X =1606 Y =1379, X =1901 Y =218, X =1226 Y =1907, X =1133 Y =795, X =72 Y =881. The V magnitudes are respectively: 9.32, 7.36, 6.82, 9.48, 6.04.

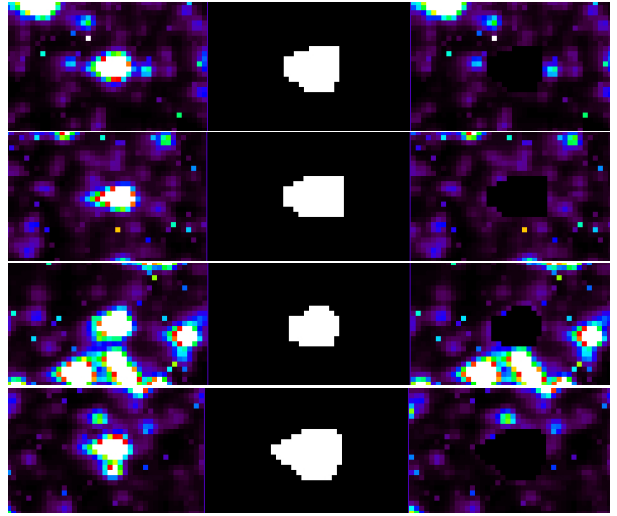

**Fig. 12.** Four PF PSF on CCD E1. From top to bottom the star position on CCDs is X = 1734 Y = 1769 , X = 67 Y = 1062, X = 434 Y = 317, X = 40 Y = 716. The stars R magnitude are respectively 12.6, 13.1, 13.0, 13.1. In each panel from left to right are shown the star image, the aperture and the field outside the aperture. See also section 5.2

#### 4.2.3. The geometric model.

The aim of the geometric model is to transform the coordinates of the stars into $(X, Y)$ coordinates in the CCD frames and vice versa. Four reference systems are used: (1) the equatorial J2000 inertial reference system is used to reference the direction of stars and the orientation of the telescope, (2) the spacecraft mechanical reference system is used to describe the incident angles of the targets, (3) the focal plane reference system references the position of the detectors in the focal plane, and (4) the 4 CCD reference systems are used to reference the coordinates of the stars on the detectors. The geometric model implements the transformations between these reference frames. The parameters of the geometric model are: the quaternion giving the orientation of the telescope, the coefficients of distortion of the optical system, the position (in mm) and orientation of the detector in the focal plane reference system, and the squared pixel size (in microns). The initial parameters of the geometric model have been derived from the optical (Zemax) model of the telescope. The parameters have been updated during the commissioning run (IRa01). The method used to update the parameters is based on a chi-square minimization between the projected coordinates of the targets in the CCD reference systems and the measured ones. In the AS channel, we iteratively calculate the telescope orientation from the positions in the AS field and the orientation of the two AS CCDs in the focal plane reference system. In the PF channel, we iteratively calculate the orientation of the two PF detectors in the focal plane reference system and optionally new values for the coefficients of distortion. At each step, the mean and maximum values of the residuals between the theoretical and measured stars positions (in pixels) are calculated. Table 2 summarises the results obtained during the first run IRa01. Figure 13 shows the position and orientation of the detectors in the asteroseismology and exoplanet fields. The maximum offset amounts to 60 $\mu$m or 4.5 pixels. The residuals amount to 0.3 pixels (mean) and 0.7 pixels (maximum) for the AS field, and 0.3 pixels (mean) and 0.9 pixels (maximum) for the PF field, without updating the coefficients of distortion. Attempts to improve the coefficients of distortion in the exoplanet fields yield a decrease of the mean residuals by less than 20 % and do not change significantly the maximum residuals. A new calibration made in the beginning of April 2008 show a translation of the PF field along the $Y$ CCD axis of 2.5 pixels or 33.7$\mu$m.

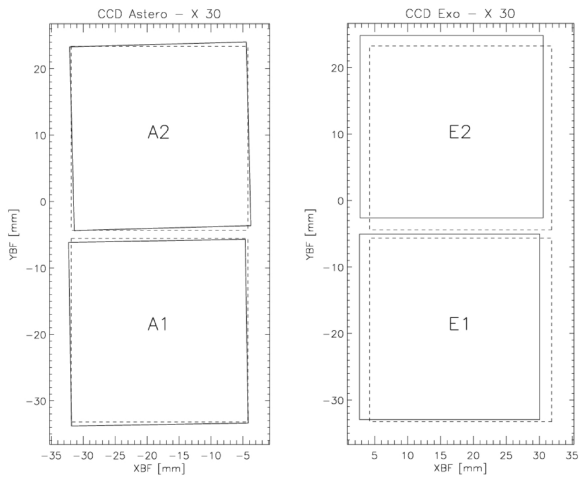

**Fig. 13.** Positions and orientations of the detectors in the AS and PF field. All the offsets have been amplified by a factor 30. The dashed line squares show the location of the detectors in the optical model. The solid line shows the location of the detectors as derived after the commissioning run IRa01. XBF and YBF is the focal bloc reference frame (unit: millimeters). The CCDs origins are given in this reference frame.

### 4.2.4. The transmission.

The transmissions of all optical subsystems were measured before launch. The transmission for the two channels is plotted in figure 14. We can see that the main effect of the lenses and prism is to cut the efficiency bandwidth on the blue side at 400 nm. The prism reduces slightly the transmission on the PF channel.

An evaluation of the global transmission in flight can be obtained by comparing measured fluxes of real stars with computed values, using model atmospheres and the transmission measured before the launch. The computation of the flux uses the Kurucz (Barban et al., 2003) atmosphere models with main parameters the effective temperature in the interval 9000 – 3500 K and a surface gravity of 4.5. After identification of the targets using Tycho2 or the Exodat catalogues (Deleuil et al. 2008), a magnitude and an effective temperature are available. For the AS programme the V magnitude is computed from the Tycho catalogue with the relation $V = V_T - 0.1284(B_T - V_T) + 0.0442(B_T - V_T)^2 - 0.015(B_T - V_T)^3$ and the effective temperature is given from the seismology data base (Charpinet et al., 2006, Solano et al., 2005). For the PF programme the V magnitude and a colour temperature are directly found in the Exodat catalogue.

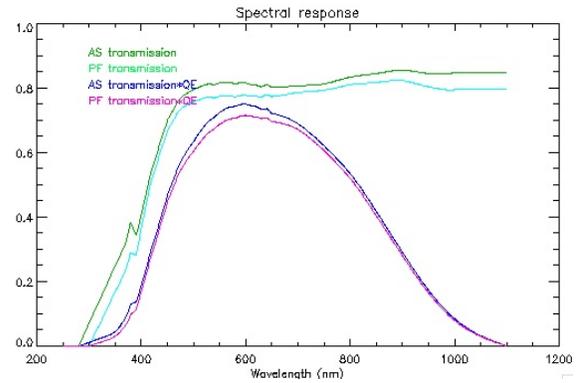

**Fig. 14.** Spectral response for the two channels. Differences between AS and PF are due to the prism absorption. The quantum efficiency (QE) has been measured for 9 wavelength values. It is the mean value for the four CCDs.

The efficiency given in figure 14 is used for photon to electron conversion.

The flux is measured on full frame images obtained at the beginning of each run. More than 120 stars are identified on each AS CCD for the three runs IRa01, LRc01 and LRa01. To minimise the measurement errors 300 bright stars with V magnitude in the interval 11.5 – 13.5, are processed in the same way on the PF field. Stars with temperatures hotter than 9000 K or cooler than 3500 K are eliminated and the relation "computed flux" versus "measured flux" is determined.

The slope is of the order of 0.94 with a dispersion of 0.05, the measured flux seems slightly larger than the computed one, this discrepancy could be due to the poor sampling of the quantum efficiency measurements (see figures 15 and 16).

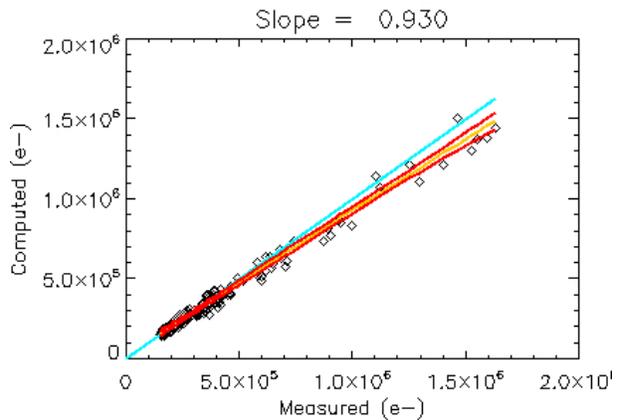

**Fig. 15.** Comparison of the computed and measured flux for the AS CCDs. Green curve: line of slope 1. Yellow curve: best polynomial fit. Red curves: one $\sigma$ error limits on the polynomial coefficients.

## 4.3. The CCD and readout chain

### 4.3.1. The CCD detectors.

The focal plane is composed of four CCDs, namely 4280 CCDs provided by E2V Technologies. These are frame-transfer, thinned, back-illuminated detectors presenting 2048 × 2048 pixels in the image zone and a memory area of 2048 × 2054 pixels, each pixel is 13.5 × 13.5$\mu$m$^2$ in size. The corresponding angular pixel size is 2.32 arcsec.

**Table 2.** In-flight position and orientation of the AS and PF detectors in the focal plane reference system, as derived during the commissioning run.

| CCD | Parameter | Value |
| --- | --- | --- |
| A1 | $X(1, 1)$ - coordinate offset of the first pixel in the focal plane reference system | −14 microns |
|    | $Y(1, 1)$ - coordinate offset of the first pixel in the focal plane reference system | −20 microns |
|    | Rotation angle in the focal plane reference system | −0.03 degrees |
| A2 | $X(1, 1)$ - coordinate offset of the first pixel in the focal plane reference system | −9 microns |
|    | $Y(1, 1)$ - coordinate offset of the first pixel in the focal plane reference system | −2 microns |
|    | Rotation angle in the focal plane reference system | −0.05 degrees |
| E1 | $X(1, 1)$ - coordinate offset of the first pixel in the focal plane reference system | −60 microns |
|    | $Y(1, 1)$ - coordinate offset of the first pixel in the focal plane reference system | −11 microns |
|    | Rotation angle in the focal plane reference system | −0.01 degrees |
| E2 | $X(1, 1)$ - coordinate offset of the first pixel in the focal plane reference system | −43 microns |
|    | $Y(1, 1)$ - coordinate offset of the first pixel in the focal plane reference system | +60 microns |
|    | Rotation angle in the focal plane reference system | −0.01 degrees |

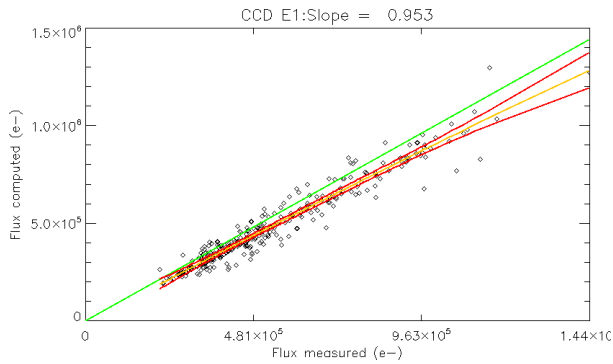

**Fig. 16.** Comparison of the computed and measured flux for the PF CCD E1. Green curve: line of slope 1. Yellow curve: best polynomial fit. Red curves: one $\sigma$ error limits on the polynomial coefficients

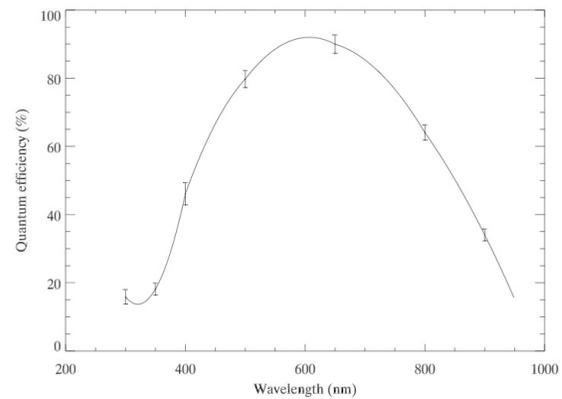

**Fig. 17.** Mean quantum efficiency of 4280 CCDs

The four CoRoT CCDs are denoted as A1, A2 when belonging to the AS programme and E1 and E2 for the PF program. Each CCD has two outputs and consequently eight electronic chains are necessary for the digitisation.

The flight detectors were selected from among 10 candidates (see Lapeyrere et al 2006 for a description of the calibration and sorting). The quantum efficiency limits the bandwidth in the wavelength interval 250 to 1000 nm. (see figure 17).

On the AS CCDs the acquisition cycle is 1 s., so there is no time for a whole image readout of 22.70 seconds, but this is possible on the PF CCDs, with a cycle of 32 seconds. In nominal observing mode only ten windows (50 × 50 pixels) can be extracted from each CCD, five star windows and five background windows binned 5×5. Additionally an over-scanned row of 2048 pixels allows the measurement of the zero level of each left and right electronic readout chain.

The exposure duration is limited by the beginning of charge transfer between the image zone to the memory zone. The row transfer on the AS CCDs lasts 100$\mu$s and 150$\mu$s on the PF CCDs. Consequently the image exposure times are respectively 0.7953 s. and 31.61928 s.

The CCDs are operated at about −36 C. giving a theoretical dark current rate of 0.6 e$^-$/pixel/s. Such a value is difficult to measure. It would require a perfectly hermetic shutter, which is not the case.

### 4.3.2. The readout electronics.

The readout electronics are responsible for the digitisation of the output signal from each half CCD. A schematic view is shown in figure 18. A pre-amplification stage (called EP) is placed in the vicinity of the CCD. The following signal processing is performed by the control camera unit (called BCC in figure 18). Digitisation is performed by a 16 bit A/D converter, for an input signal ranging from 0 to 5 V which corresponds to 76.3 $\mu$V/ADU. The CCD capacity being 120000 e$^-$ the total gain is about 2 e$^-$/ADU

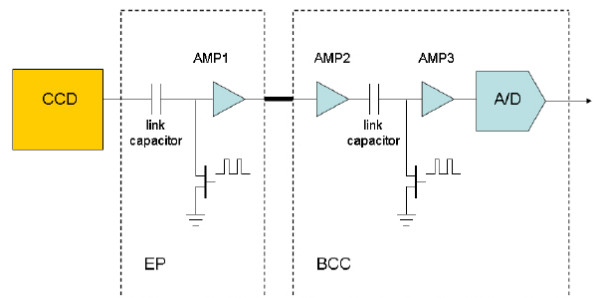

**Fig. 18.** Schematic of the readout electronics.

The estimate of the gain exploits the fact that if the statistical distribution of detected photons has a Poisson distribution, it is not preserved after readout when it is passed through a non unitary gain block. (Janesick, 2001, Pinheiro da Silva et al., 2006). We compute the relation mean-variance for images of different intensity levels. The proportionality factor represents the gain value itself. To avoid the pixel response non uniformity contribution and low frequency non uniform contributions to the variance, it is computed on an image which is the difference of two independent images with the same mean intensity level (see figure 19 and table 3 for the results). In this case the proportionality factor is twice the gain.

The readout noise is measured through the readout of a pre-emptied CCD output register. This line, called the offset line, is registered at each exposure on the eight half-CCDs (table 3).

nised, which means that along a run the cross-talk perturbations always affect the same pixels and that the perturbation intensity is stable on a long time scale (> 32 s). 15 different low level commands are used for the charge transfer and pixel digitisation. The most important sequences are: Pixel Digitisation 10 $\mu$s, line transfer in the register 150 and 250 $\mu$s, binning in the summing well 20 $\mu$s, clear register 100 $\mu$s, line transfer from the image zone to the memory zone 100 and 150 $\mu$s. Knowing which sequences are active at each time step we can compute the location of the perturbation, and its intensity is calibrated for each pixel on actual images.

Figure 20 shows an example of cross talk on the PF CCD due to the AS readout, and figure 21 the perturbation during a 32 seconds cycle of an AS 50 × 50 pixels image due the PF CCD readout.

Figure 22 shows the intensity of the perturbation on an AS CCD due to the PF line transfer in the register command.

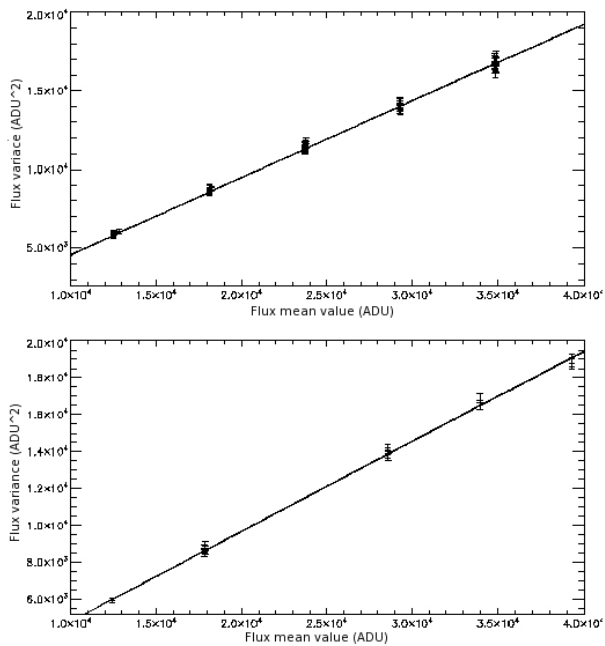

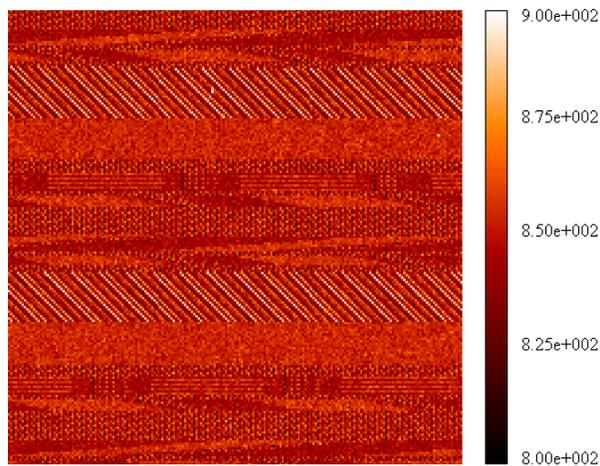

**Fig. 20.** Perturbation pattern on a 150 × 150 pixel sub-image of a PF CCD.

**Fig. 19.** Mean-Variance relation for two CCDs. Top A1 , bottom E1.

**Table 3.** Readout noise and gain. L and R stand for left and right CCDs output

| CCD name | A1L | A1R | A2L | A2R |
|---|---|---|---|---|
| Readout noise (ADU/pixel/readout) | 4.34 | 4.26 | 4.25 | 4.36 |
| Gain ($e^-/ADU$) | 2.04 | 2.05 | 1.99 | 1.99 |
| CCD name | E1L | E1R | E2L | E2R |
| Readout noise (ADU/pixel/readout) | 5.5 | 5.4 | 5.5 | 5.3 |
| Gain ($e^-/ADU$) | 2.06 | 2.05 | 2.15 | 2.14 |

All stars show a long term decrease of the flux roughly linear in time. The slope, measured on the 146 days of the run LRc01 for the ten AS stars, varies linearly with the absolute flux value. Attributing the flux decrease to a gain decrease, the gain variation is close to $-1.\,10^{-4}$ days$^{-1}$.

### 4.3.3. Cross talk perturbations

The main cross-talk perturbations are between the PF and AS CCDs. As we already said all readout processes are synchro-

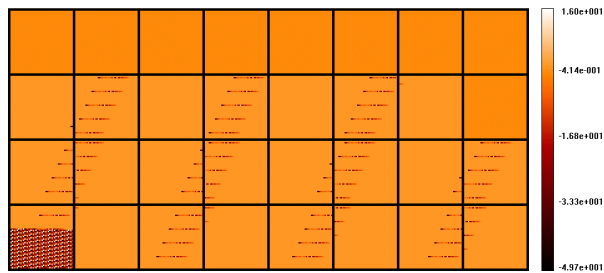

**Fig. 21.** Perturbation on a 50×50 pixel AS window during a 32 seconds cycle. The first exposure is the bottom left image and the last exposure is the upper right image. The first exposure is strongly perturbed by the end of the image zone to the memory zone charge transfer of the PF CCD. The nine last exposures are free of perturbations, the PF readout lasting only 22.7 seconds.

### 4.4. Thermal control.

To ensure the stability of the photometric signal, the main thermal requirements are:

- A telescope stability better than 0.3 K on the orbital time scale and lower than 1 K on a 150 days run.

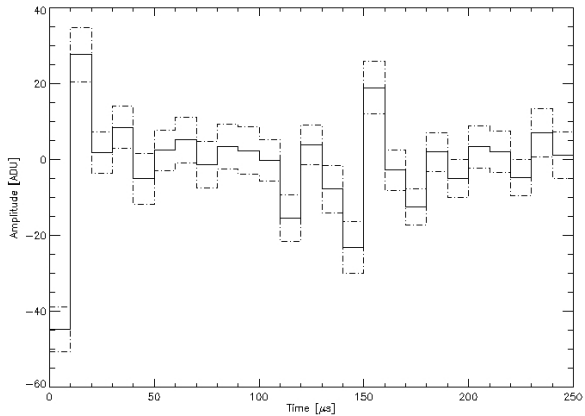

**Fig. 22.** Example of the perturbation on an AS CCD by the" line transfer in register" command of a PF CCD. The command lasts 250 $\mu$s and the perturbation is calibrated in 10 $\mu$s steps. The amplitude can reach 80 ADU. The dashed lines show the 1 $\sigma$ uncertainty.

- A detector temperature lower than −40 C and a stability of 0.02 C peak to peak on the orbit.
- A stability of the readout electronics better than 0.3 C peak to peak on the orbit.

The thermal structure is made of five isolated thermal enclosures, the external baffle, telescope, upper equipment bay including the CCD readout electronics, lower equipment bay, and the focal plane assembly. In order to reduce as much as possible the orbital variation a systematic increase of passive thermal inertia is used by increasing the components mass. For an extensive description of the thermal design see Hustaix et al., 2007

Ten PROTEUS thermal control heaters are used by the CoRoT payload. More specifically, there are two regulation lines in the vicinity of the mirror M2, one for the mirror M1, two lines on the dioptric objective, one on the focal plane assembly, two on the honeycomb plateau supporting the mirrors and two on the equipment bay. A specific and very accurate heater line provides the detectors stability to smooth out the radiator's orbital and seasonal variations (figure 23). This heater line uses sensitive and linear sensors (AD590) and a pure proportional method using coefficients which give a quick response (8 Hz) of the system. As the radiator temperature varies along a run between −60 and −80 C the target regulation temperature must be changed several times to ensure the CCDs orbital fluctuations are less than 0.01 C peak to peak. Consequently the mean CCDs temperature varies from −40 to −36 degrees.

The correlation between the photometric signal and temperature variations is difficult to determine, particularly on an orbital time scale, as several subsystems can experience orbital temperature perturbations and different physical processes may affect the LC on the same time scale. We have therefore chosen to characterise only the effect of the CCDs temperature on the LC, when the mean target temperature is changed. The temperature variation looks like a "discontinuity" and is easier to characterise in the offset, background and stars LCs.

The relation between the CCD temperature and the offset is different for the eight readout chains. For instance the offset value decreases on the AS CCDs but increases on the PF CCDs. On the AS channel the flux variations are very small and difficult to measure. Figure 24 shows the LC decrease for a background LC on an AS CCD and averaged (see section 7) and normalised PF light curves. The coefficient relating the CCD temperature variation to the LC is of the order of 0.22 e$^-$pixel$^{-1}$C$^{-1}$ for the background and 10 e$^-$pixel$^{-1}$C$^{-1}$ for a m$_R \sim$ 13.5 star. We have not found a unique relation between the stars flux and the CCDs temperature.

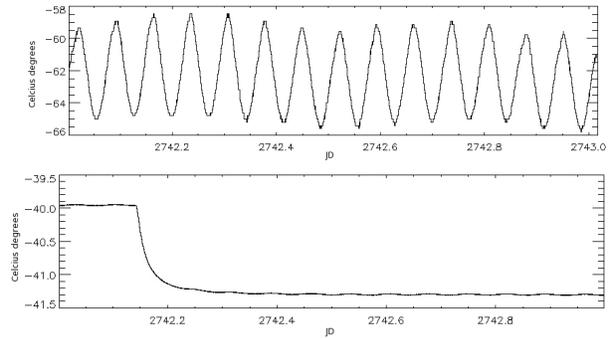

**Fig. 23.** Upper panel: radiator temperature during 24 hours. The orbital amplitude is 6 degrees. Lower panel: temperature of the CCDs around a target temperature jump (27$^{th}$ july 2007). The mean temperature jump is 1.4 degrees.

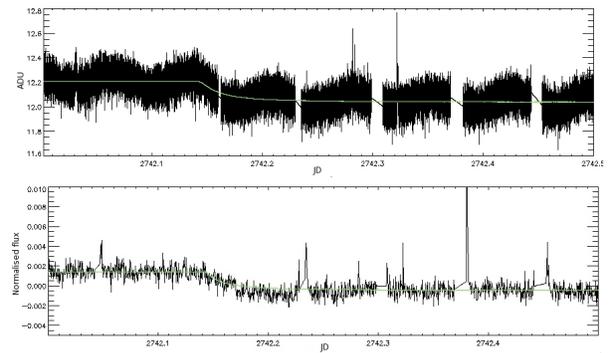

**Fig. 24.** From top to bottom: Background LC variation on an AS CCD. Averaged and normalised PF LC for stars with m$_R$ in the interval 13−14. The green curves are the best fit to the CCDs' temperature variations.

### 4.5. Digital assembly and onboard software.

#### 4.5.1. Digital assembly hardware.

The digital assembly is made of two chains managing the data from one AS CCD and one PF CCD. The two chains are identical. In each chain there are two main functions: to select the useful pixels to be processed (Extractor Box or BEX) and to process the data (Digital Processing Unit or DPU). The BEX receives all pixels digitised by the readout electronics and group pixels belonging to a star into a specific buffer. The BEX can manage different image shapes: full frame image, rectangular sub-image of various size, or pixels included in a photometric aperture.

#### 4.5.2. On board software.

This is divided in two parts: the Primary Boot Software (PBS) and the APplication Software (APS) located in an EEPROM or RAM. The PBS is active when the DPU is ON and the APS starts on a tele-command (TC) send from ground. The main PBS functions are:

- Initialise the DPU after power-on
- Collect and provide status telemetry data
- Receive and execute low level telecommands

- Pack telemetry in the buffer and transfer it to the PROTEUS mass memory
- Start and maintain the APS executables
- Check health and handle error events

The APS is the set of tasks which extracts the photometric signal from the images for the two channels, receives and interprets the telecommands, and packs and transfers the "scientific data". The APS processes each second two offset, five background and five star windows, and each 32 seconds a maximum number of 6000 objects ( 5720 star, 200 background and 80 offset). The telemetry content is summarised in table 4. The BEX, the DPU and the onboard software work exactly as expected.

### 4.6. On-board time.

To synchronise all the readout processes a stabilised quartz of 10 MHz inside the payload produces all signals. The basic time scales are $10\mu$Hz for the pixel digitisation. and 100 $\mu$Hz and 150 $\mu$Hz for the line transfer from the image to memory zone respectively on the AS and PF CCD. A signal sent by the PROTEUS clock, named Real Time Clock (RTC), each 32 seconds, starts the PF CCD image to memory transfer. On the AS CCD, in a cycle of 32 seconds, the first 31 transfers start on a one second signal given by the payload quartz and the last transfer starts on the RTC signal. There is no synchronisation between the payload quartz and the RTC.

On PROTEUS a counter measures the RTC quartz frequency. A GPS receiver delivers each second a UT date and a one second signal, the stability of which is 1 $\mu$s. When the counter receives this signal its value is registered and written in the housekeeping telemetry. With the UT date and the counter values we transform the on board time to UT and correct the RTC frequency fluctuations due essentially to the temperature variations of the quartz.

### 4.7. The Satellite Attitude Control System (ACS).

The attitude control system hardware is composed of a star-tracker, gyroscopes, inertial wheels and magnetotorquers. The wheels desaturation is performed by the magnetotorquers in a continuous way to avoid strong perturbations of the line of sight, this is done several time each orbit. The inertial wheels and magnetotorquer command laws are determined by a Kalman filter which receives velocity information from the star-tracker and from the gyroscopes. The stability of the standard PROTEUS ACS is 16 arc-seconds, far from the CoRoT specification of 0.5 arc-second. To reach the specification during the observing phases, the instrument replaces the star-tracker. With two bright stars on one of the two AS CCD the pointing error is computed each second and sent to the Kalman filter with a delay of 1.650 seconds. With this improvement the rms pointing stability is of 0.15 arc-second on the transverse axis and 5 arc-seconds around the line of sight. This third angle gives a negligible contribution to the field motion on the CCD. At eclipse transition (light to shadow or shadow to light) rapid temperature variations of the solar arrays (from -80 to +70 C) produce vibrations and consequently attitude fluctuations twice each orbit. The amplitude can reach 12 arc-second depending of the arrays' orientation and the position of the Sun. The perturbation of the stellar light curve typically lasts 30 seconds with a delay of the order of 10 seconds at the light to shadow transition, and 50 seconds at the eclipse egress. An example of eclipse perturbation is given in figure 25 for a AS ($m_v$ = 6.74) star.

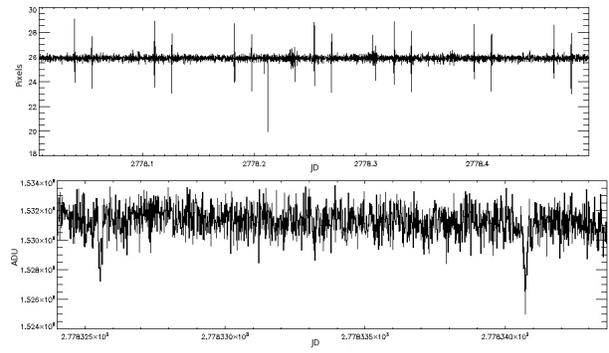

**Fig. 25.** Upper panel: baricentre X with periodic solar array perturbations. Lower panel: detail of the light curve on an eclipse ingress and egress

The amplitude of the pointing perturbation of the LC flux depends on the aperture, PSF shape and on the depointing amplitude. It is generally of the order or less than 0.4 % but can reach 20 % for eclipse ingress/egress.

The pointing error is computed with the measured position of two bright stars on one AS CCD. To obtain the best precision on the baricentre, the offset and background measured on board are first subtracted from the stellar image. Then pixels brighter than a threshold (typical value 400 ADU) are selected and used to compute the baricentre. With such an algorithm the position measurement is sensitive to proton impacts, but this is not a real problem as during the SAA crossing the photometric signal is also strongly perturbed independently of the pointing fluctuations.

## 5. Extracting the photometric signal.

To comply with the telemetry volume available (1.5 Gbits/24 hours) the photometric signal must be extracted on board for most of the targets. An aperture photometry method is used. The problem is to define the pixels belonging to the aperture giving a flux measurement which maximises a signal to noise ratio. While it is possible to keep in memory five arrays describing the aperture in the AS channel, it is not possible for the 6000 PF objects. The number of PF apertures being limited to 255 on board, the problem is to reduce 6000 optimised apertures to 255 without a significant decrease of the signal to noise ratio.

### 5.1. Aperture on the AS channel.

As only ten stars are processed on the AS CCDs it is possible to define for each star an aperture and upload it on board. The signal to noise ratio is defined in a classical form as:

$$\frac{S}{N}(k) = \frac{\sum_{i=1}^{k} N_e(i)}{\sqrt{\sum_{i=1}^{k} \left(N_e(i) + \sigma_i^2\right)}} \quad (4)$$

$N_e$ is the pixel intensity sorted by decreasing order. The maximum value of $S/N$ gives the set of pixels belonging to the aperture. The total noise can be defined in this simple form if its contribution is statistically independent in each pixel, but this hypothesis fails with jitter noise. All pixels on the aperture edges contribute to the jitter noise. An iterative method (Fialho and Auvergne, 2006) is used to define the aperture using as initial condition an aperture deduced from equation 4. The algorithm steps are:

**Table 4.** Main raw data for the two programmes with time sampling. Note that on the AS channel the offset and background are subtracted on board.

| Variable | programme | Sampling (sec.) | Comment |
| --- | --- | --- | --- |
| Mean offset | AS | 1 | Averaged on 1024 pixels, (ADU px$^{-1}$) |
| Bright pixel number | AS | 1 | Number of pixels brighter than $n\sigma$ |
| Offset variance | AS | 1 | Readout noise, (ADU px$^{-1}$) |
| Star Intensity | AS | 1 | Total inside the aperture, (ADU) |
| Baricentre X | AS | 1 | star X position, (px) |
| Baricentre Y | AS | 1 | star Y position, (px) |
| Background Intensity | AS | 1 | Pixels brighter than $n\sigma$ are not included in the mean, ADU px$^{-1}$ |
| Bright pixel number in background window | AS | 1 | Number of pixels brighter than $n\sigma$ |
| Offset | PF | 32 | mean offset , ADU px$^{-1}$ |
| Background | PF | 32 and 512 | mean background, ADU px$^{-1}$ |
| Star white flux | PF | 32 and 512 | star total intensity in the aperture, ADU |
| Star "red" flux | PF | 32 and 512 | star total intensity in the red part of the aperture, ADU |
| Star "green" flux | PF | 32 and 512 | star total intensity in the green part of the aperture, ADU |
| Star "blue" flux | PF | 32 and 512 | star total intensity in the blue part of the aperture, ADU |

- PSF estimation of a super resolved PSF
- PSF fitting to separate target and parasite star. If the parasite shares pixels with the target those pixels are included in the noise contribution of the signal to noise ratio.
- Computation of the $S/N$ with equation 4. With its maximum value an aperture is determined.
- With this aperture the photometric signal is computed on a sample of images specially downloaded for this task. Then the photometric signal is corrected with the method described in section 6 and a $S/N$ is computed.
- An edge pixel is added to the aperture and a new $S/N$ computed from the fourth step.
- The process stops when the maximum $S/N$ is reached.

### 5.2. Aperture on the PF channel.

A detailed description of the method can be found in Llebaria et al. 2002 and in Llebaria et al. 2004 and an evaluation of the signal to noise obtained in Llebaria et al., 2003.

The star fields to be observed are sometimes quite crowded and the aperture shape will depend on: 1) the PSF distribution, 2) the local background and 3) the neighbouring stars. The best aperture for each star should minimise such effects to approach the ideal S/N ratio of photon limited noise. The signal is defined by the total flux of the star included in the aperture. The noise depends mainly on:

- photon noise from the star itself
- photon noise from the background and the overlapping stars
- residual jitter
- focal length orbital variations (negligible)
- periodic straylight variations

An initial aperture is optimised for each star assuming zero jitter noise. In the second step taking into account jitter, apertures are modified to give minimal variance relative to the stellar flux. The third step is to reduce the full set of apertures from all selected stars to the limited set of 256 shapes. In the last step an aperture is assigned to each star. In the worst case the final $S/N$ is reduced by 20% compared to a $S/N$ ratio limited only by the photon noise.

Four examples of aperture assignment can be seen on figure 12. From top to bottom, the two first panels show the aperture of an isolated star. The third one shows that the aperture is smaller than for the second case, and placed in such a way that parasite stars just below are not included in the aperture. The fourth panel shows a parasite star inside the target star. In that case the aperture is large and includes the parasite.

## 6. Outline of data correction.

The main processing and corrections on the raw photometric signal are roughly the same for the two programmes (Samadi et al., 2006).

- Correction of the offset and background signal from the cross-talk effect
- Subtraction of the offset and background to the star light curve
- Stellar flux correction of the cross-talk effect (AS only)
- Stellar flux unit transform from ADU to electrons
- Correction of the exposure time variations
- Correction of the jitter effect
- Correction of the outliers.

### 6.1. Background corrections on the PF channel.

The background is measured on 400 windows of $10 \times 10$ pixels in the PF field, 300 with a sampling of 512 seconds and 100 with 32 second sampling; they are uniformly spread over the field in regions free of stars. The issue is how to estimate the background inside a star aperture. Four methods have been tested (Drummond et al., 2008) they are: closest background, triangular interpolation, two dimensional polynomial fit and the median of all background windows. The last two methods are the most efficient and are insensitive to bright pixels but the median is overall the most effective to correct orbital background variations, despite the fact that all the stars LCs on one CCD are corrected with the same background. The bright pixel problem in background windows has been solved recently with a new version of the on board software uploaded in April 2008, the brightest pixels inside a background window image are removed before the computation of the mean.

### 6.2. Jitter effect corrections.

The jitter corrections are completed in five steps (for a detailed description of the method see Fialho et al., 2007 and Drummond et al., 2006):

- compute for all stars a high resolution PSF (1/4 of the pixels for AS channel and 1/2 the pixels for PF)

- compute a least square solution of the line of sight fluctuations with stars on the AS CCDs brighter than V=8.
- compute the motion of all stars in the CCD reference frame from the angles obtained in the second step.
- moving on a regular sampled grid the high resolution PSF inside the aperture, the flux
- transferring the high resolution PSF inside the aperture onto a uniform grid, the flux (normalised to unity) is computed for each grid point. This surface, computed for each target, gives the proportion of flux lost when the image moves outside the aperture.
- for actual motions the surface generated in the fourth step is interpolated to compute a correction factor applied to the measured flux.

Figure 26 shows three spectra of a faint star on the AS channel. The upper one is the spectrum obtained before the jitter correction. A lot of lines, harmonics of the orbital frequency appear with a continuous component. This continuous component is centred at a frequency around 0.03 Hz (or 35 seconds). After correction most of the lines and the continuous component disappear. The third spectrum is computed with data obtained in July 2007, a period without eclipses, and before jitter corrections. Comparing the spectra of August and July we can see that most of the lines come from eclipse ingress/egress and that a small residual remains. The efficiency of the correction depends strongly on the high resolution PSF estimation quality. Generally the PSF used to compute the correction surface is precise enough, particularly for bright stars, but sometimes for faintest stars the PSF is not precise enough in the PSF wings. The example shown in figure 26 is that of a worst case.

This algorithm has not been implemented yet on the PF light curves, the PSF computation being more difficult than for the AS objects, in part because the integration time is longer but also because the objects are faint and their images undersampled. But some tests have been made of the proposed method and examples of the results are shown in figure 27 for two stars, a binary star and a constant one. On this second star we can see a small correction residual but the perturbation is reduced by more than a factor 40 (Fialho, 2008).

### 6.2.1. Outliers.

Outliers are data points in a star's LC which are significantly larger than the average flux. Sometimes the reason of the perturbation is known or suspected (image of debris, high energy particles) but this is not often the case. Nevertheless, outliers are removed assuming that very short photon excess can not be due to stellar flux fluctuations. The outlier detection is based on a robust and local standard deviation. Flux values which differ by more than $n\sigma$ from the mean are removed and replaced by an interpolated value.

## 7. Photometric performances.

The measurement of the LC rms is carried out in Fourier space on frequencies adapted to each scientific programme. For the AS programme the noise level is measured in the frequency interval 1 to 10 mHz, which corresponds to the range of solar-like oscillations. In the PF channel the frequency interval is $2.5\ 10^{-5}$ to $1.5\ 10^{-4}$ Hz which corresponds to a typical transit duration. For the relation between the spectrum mean value and the rms see for instance Bracewell,1986.

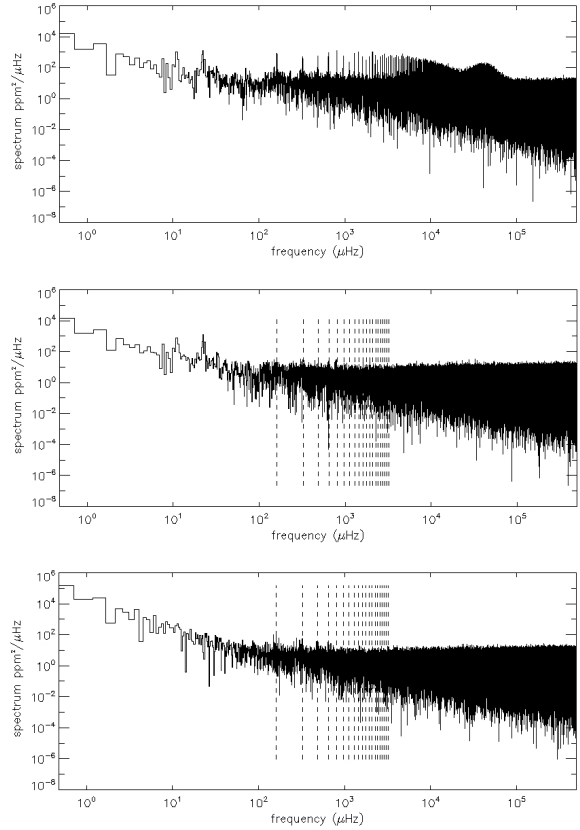

**Fig. 26.** From top to bottom: 1) Power spectrum of a constant faint (V = 9.14) star in the AS channel for 15 days of data (August 2007) with eclipses, but before jitter correction. 2) Spectrum of the same data after correction. 3) Spectrum of the same star for 15 days data (July 2007) without eclipses and before jitter corrections. Dashed vertical lines indicate the orbital frequency and its harmonics.

### 7.1. Photometric noise at high frequencies on AS channel.

The spectrum is computed using 145 days of the run LRc01. The data are corrected as described in section 6 and the results shown in figure 28. For comparison with the specification of 0.6 ppm in 5 days (for a V = 5.7 star) the result is normalised to correspond to a five day run. The equivalent photon noise is given for comparison. The small differences between the photon noise and the real one, around V = 5.8, is due to the jitter correction residual. A recent paper on the star HD 49933 (Appourchaux et al., 2008) reports the detection of several tens of solar-like p-mode oscillations with a signal to noise ratio down to 1.5 and amplitude less than 0.6 ppm/$\sqrt{\mu Hz}$.

### 7.2. Photometric noise at intermediate frequencies on the PF channel.

Before computing the Fourier transform, a "mean LC" is computed in four magnitude intervals. The operations performed are: select stars in a magnitude interval, normalise the mean flux to 1 and eliminate variable stars, perform a Singular Value Analysis and reconstruct the mean LC on the subspace corresponding to the largest eigenvalue. The result is given in figure 29.

However the best performance evaluation is given by transit detection. Figure 30 shows all the transits detected in the first three runs (Barge, 2008). More than 400 events are reported. All the plane accessible to CoRoT is filled. The smallest tran-sit depth is around $6\,10^{-4}$ (found in a star of magnitude R = 12), and is compatible with measurements shown in figure 29. For short or intermediate periods a large number of transits allows a detection close to the theoretical limit. The longest periods are of the order of 75 days, which is the longest period for which we can have three transits as required by the detection algorithms.

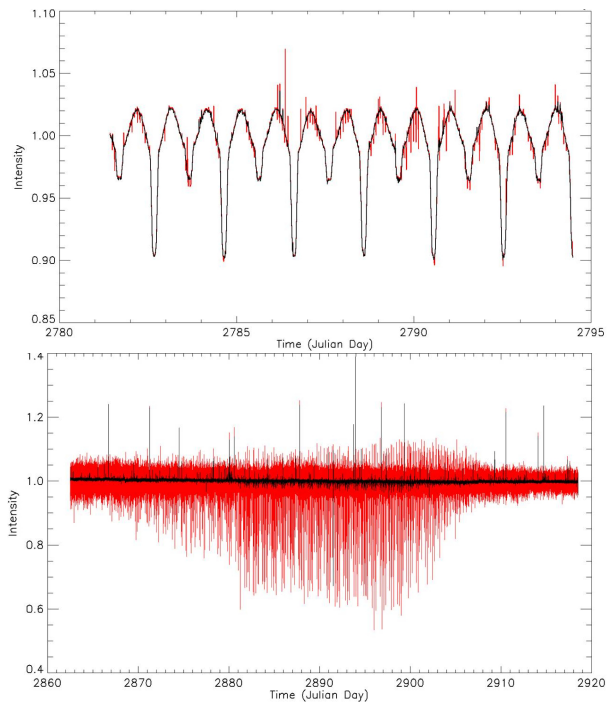

**Fig. 27.** Red light curve on the PF channel for two stars. Upper panel: a binary star sampled at 512 s ($m_R = 13.7$). Lower panel: a constant star sampled at 32 s ($m_R = 14$). Red curves: raw data before correction. Black curves: the signal after correction. The flux is normalised.

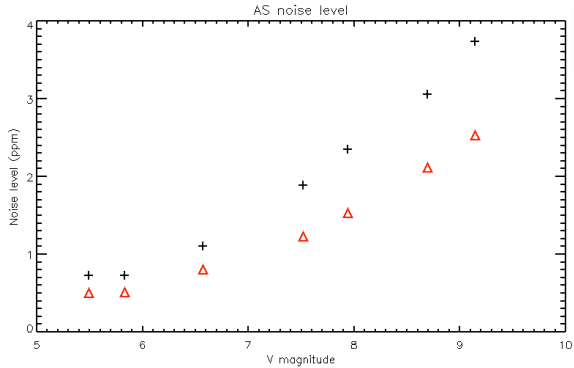

**Fig. 28.** Crosses: measured rms. Red triangles: photons noise level.

## 8. Conclusions.

We have analysed the data of the commissioning and of four following runs. Most of the subsystems have nominal performances:

- The scattered light from the Earth is reduced by a factor of the order of $10^{12}$ by the baffle, the residual background orbital variations being always smaller than 0.6 $e^-s^{-1}pixel^{-1}$. The specification asks for a variation less than 5 $e^-s^{-1}pixel^{-1}$

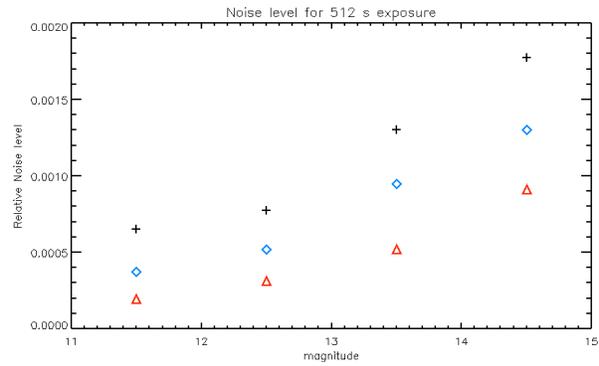

**Fig. 29.** Crosses: measured rms on white light curve . Blue diamond: the same but measured on a time interval without eclipses. Red triangles: white noise level (photons noise plus readout noise plus background photons noise)

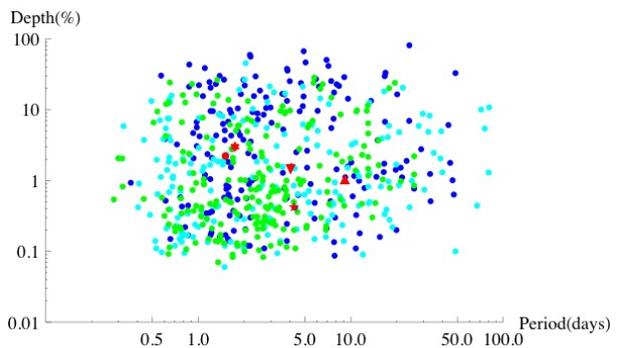

**Fig. 30.** Transits detected in CoRoT data during the three first runs (dark blue, light blue and green points). Confirmed planets are red points.

- The PSF is close to the expected one, despite a slightly larger defocus. The photon detection efficiency is in agreement with calibration and transmission measurements.
- The temperature stability on short time scales is always better than the specified performances. On the CCDs the stability specification for the orbital variations was 0.03 degree the actual stability is better than 0.01
- The ACS gives a stability of the line of sight of 0.15 arcsecond, much better than the specification of 0.5 arcsecond. While we did not foresee the effect of eclipse ingress/egress on the ACS this perturbation can be efficiently corrected.

We have shown that most of the photometric perturbations come from the environmental perturbations and especially from eclipse ingress/egress. Hot pixels generated by proton impact can not be easily corrected in stellar light curves, particularly for variable stars, but hot pixels brighter than 1000 $e^-$ are rare enough ($\sim 1\%$) to be acceptable. The last puzzling perturbations are the temperature fluctuations which introduce variations on 24 hours and orbital time scales. It is not yet clear whether we will be able to obtain a unique calibration for all stars.

[1] Observatoire de Paris, UMR 8109, 5 place J. Janssen, 92195 Meudon, France
[2] Centre National d'Etudes Spatiales. 18 avenue E. Belin, Toulouse, France.
[3] Laboratoire dAstrophysique de Marseille, Traverse du Siphon, 13376 Marseille cedex 12, France
[4] Institute of Planetary Research, DLR, Rutherfordstr. 2, D-12489 Berlin, Germany
[5] Institut dAstrophysique Spatiale, Université Paris XI, F-91405 Orsay, France
[6] DLR, German Aerospace Center, Institute for Robotics and Mechatronics, Department Optical Information Systems, Rutherford Strasse 2, D-12489 Berlin-Adlershof
[7] Technologiepark 1, D-15236 Frankfurt/Oder
[8] CLIPhIT, Arnfriedstrasse 17, D-12683 Berlin
[9] European Space Agency, ESTEC, SCI-A, P.O. Box 299, NL-2200AG, Noordwijk, The Netherlands.
[10] Space Research Institute, Austrian Academy of Science, Schmiedlstrasse 6, 8042, Graz.
[11] Centre Spatial de Liége, ULG Science Park, Av. du Pré-Aly 4031 Angleur-Liége.
[12] Observatoire Midi-Pyrénées, 14 Av. E. Belin, Toulouse, France
[13] Instituut voor Sterrenkunde, Departement Natuurkunde en Sterrenkunde , Katholieke Universiteit Leuven , Celestijnenlaan 200 D, Office 03.31 , B - 3001 Leuven